\documentclass[sigconf]{acmart} 

\usepackage[utf8]{inputenc}
\usepackage{url}
\usepackage{xspace} 
\usepackage{soul}
\usepackage{multirow}

\copyrightyear{2018}
\acmYear{2018}
\setcopyright{acmlicensed}
\acmConference[SEEM'18]{SEEM'18:IEEE/ACM International Workshop on Software Engineering Education for Millennials }{May 27-June 3 2018}{Gothenburg, Sweden}
\acmBooktitle{SEEM'18: SEEM'18:IEEE/ACM International Workshop on Software Engineering Education for Millennials , May 27-June 3 2018, Gothenburg, Sweden}
\acmPrice{15.00}
\acmDOI{10.1145/3194779.3194787}
\acmISBN{978-1-4503-5750-0/18/05}

\usepackage{ifthen}

\newboolean{showcomments}
\setboolean{showcomments}{true}
\ifthenelse{\boolean{showcomments}}{%
   \newcommand{\bnote}[2]{
    \marginpar{\fbox{\bfseries\sffamily#1}}
    \fbox{\bfseries\sffamily\scriptsize #1}
    {\sffamily\small$\blacktriangleright$#2$\blacktriangleleft$}}
}{
   \newcommand{\bnote}[2]{}
}

\begin{document}

\title{Software Engineering for Millennials, by Millennials} 
\titlenote{We picked this title because the teaching staff for 2017 are all Millennials. }

\author{Jocelyn Simmonds}
\affiliation{%
  \institution{Department of Computer Science,\\ University of Chile}
}
\email{jsimmond@dcc.uchile.cl}

\author{Maíra Marques Samary}
\affiliation{%
  \institution{Department of Computer Science,\\ University of Chile}
}
\email{mmarques@dcc.uchile.cl}

\author{Milenko Tomic}
\affiliation{%
  \institution{Department of Computer Science,\\ University of Chile}
}
\email{mtomic@dcc.uchile.cl}

\author{Francisco Madrid}
\affiliation{%
  \institution{Department of Computer Science,\\ University of Chile}
}
\email{fmadrid@dcc.uchile.cl}

\author{Constanza Escobar}
\affiliation{%
  \institution{Department of Computer Science,\\ University of Chile}
}
\email{cescobar@dcc.uchile.cl}

\renewcommand{\shortauthors}{J. Simmonds et al.}

\begin{abstract}
Software engineers need to manage both technical and professional skills in order to be successful. Our university offers a 5.5 year program that mixes computer science, software and computer engineering, where the first two years are mostly math and physics courses. As such, our students' first real teamwork experience is during the introductory SE course, where they modify open source projects in groups of 6-8. However, students have problems working in such large teams, and feel that the course material and project are ``disconnected''. We decided to redesign this course in 2017, trying to achieve a balance between theory and practice, and technical and professional skills, with a maximum course workload of 150 hrs per semester. We share our experience in this paper, discussing the strategies we used to improve teamwork and help students learn new technologies in a more autonomous manner. We also discuss what we learned from the two times we taught the new course. 
\end{abstract}

\keywords{Project and problem-based learning, peer evaluation, gender}

\ccsdesc[500]{Social and professional topics~Computer science education}

\maketitle

\section{Introduction}

Successful software engineers must be technically competent, but must also effectively communicate with clients and users, work in teams, design and develop solutions, etc. The \emph{ACM Curriculum Guidelines for Undergraduate Degree Programs in Software Engineering} (SE2014)~\cite{SE2014} indicate that graduates of undergraduate software engineering (SE) programs should demonstrate a mastery of SE and technical knowledge, designing solutions tailored to the project context, evaluating trade-offs, while being effective team members and leaders. Chilean educational institutions offer two general computer-related programs: Computer Science Engineering\footnote{In Spanish: \emph{Ingenier\'{\i}a Civil en Computaci\'on} and/or \emph{Inform\'atica}.} (CS Engineering) and Computer Science Technology\footnote{In Spanish: \emph{Ingenier\'{\i}a en Ejecuci\'on con menci\'on en Computaci\'on} and/or \emph{Inform\'atica}.} (CS Technology). With a duration of 5-6 years, these programs start with 2 years of physics and mathematics, and then cover the fundamentals of computer science, computer and software engineering. Marques et al.~\cite{Marques-ITICSE2016} studied the CS Engineering and Technology programs offered by 20 Chilean universities, concluding that these programs only include an average of 3.4 and 2.8 SE related courses, respectively, representing only 7.4\% and 9.3\% of the courses covered by these programs. 

As such, Chilean students encounter their first SE course around their 4th year. Students at our university have 3 mandatory SE courses: CC4401 - Software Engineering I (4th year), CC5401 - Software Engineering II (5th year), and CC5402 - Software Project (5th year). Both CC4401 and CC5401 have a regular workload (150 hrs of expected dedication over 15 weeks). CC5402 is a capstone course and has double workload. Prior to taking CC4401, students have worked on individual coding assignments, requiring at most 500 lines of code (LOC), in Java, Python or C.  Students have a varied level of experience with frameworks and development tools. They rarely engage in teamwork, preferring to work alone and not letting others' (lack of) work affect their grades, because of the  competitive environment at the University of Chile, which is consistently one of the top ranked engineering schools in Chile. 

In this paper, we discuss how we redesigned the CC4401 course in order to achieve a better balance between technical and professional skills. The original course had students working in teams of 6-8 on maintaining an open source desktop application (1-10 million LOC in Java, like the BlueJ Java IDE\footnote{\url{https://www.bluej.org/}}). Teams needed to propose the change requests for their projects, following Rajlich's~\cite{Rajlich:2011} maintenance and evolution process. There were three 4-week iterations, each iteration ended with a demo of the software. Students also attended two 1.5 hr lectures per week, sat for two mandatory written evaluations and an optional exam. 

Given their lack of teamwork and development experience, most students felt ``lost'' with such a big project, and often sliced it up to avoid teamwork. We decided to redesign CC4401 in order to achieve a better balance between technical and professional skills. We also have to cover SE fundamentals and give students enough time to work on their projects. Another reason behind this decision was student motivation. Students want more time to work on their own projects, and are quick to express their frustration with what they consider ``busywork'', assignments which seem disconnected to real world problems. Given the low amount of hours dedicated to SE courses in Chile, we believe that other universities face similar challenges in their programs. 

The new version of CC4401 has a similar evaluation structure as the previous version (project, written evaluations, exam). The course project has changed: students now create a web application from scratch using Django\footnote{\url{https://www.djangoproject.com/}}. We picked Django because it has a soft learning curve. We also incorporated more partial evaluations in order to reinforce professional skills (report, presentation, peer-evaluation). The teaching staff designs the project, picking a problem that should appeal to most students, like building an application that will let them to buy lunch on campus in a quicker fashion. Students begin working in teams of 3-4 people, with the project divided up into 4 smaller, more focused iterations (requirements, user interfaces, development, and evolution \& testing). Students vote for the user interfaces that will be developed during the third iteration, as well as the code base for the fourth iteration. Our goal with this strategy was to make students read and modify code written by others, one of the key aspects of the original course project. 

We wanted to understand if the changes to CC4401 made a difference in student grades, attitudes and motivation. In order to avoid conflating changes to the course design with changes to the teaching staff, we present the data from the three semesters where the main author of this paper taught the course:  the 2nd semester\footnote{Semesters last 15 weeks, with the 1st semester spanning March-June, and the 2nd one August-December} of 2014 (original version), and the 1st and 2nd semester of 2017 (new version). The main difference between the two 2017 semesters is that students created the teams during the 1st semester, and we used a diagnostic test to set up teams during the 2nd semester. We also checked whether this strategy had an impact on student grades: it did not. Also, we found that switching students between teams did not affect their grades. Finally, we studied student peer evaluations, finding that female students were better evaluated than their male teammates, in general. We also discuss teaching evaluations for 2014 and 2017, students seem more motivated by the new course. 

This paper is structured as follows. We discuss related work in Sect.~\ref{sec:related}. In Sect.~\ref{sec:intervention}, we describe the original CC4401 course and the issues that we detected, and then describe the new version and how it addresses the previously identified issues. We then talk about our experience teaching both versions of the course in Sect.~\ref{sec:data}. We discus the lessons we learned from this experience in Sect.~\ref{sec:lessons}.

\section{Related Work}
\label{sec:related}

Before describing our experience redesigning the CC4401 course, we first discuss related work. We first give an overview of how others have introduced and enforced teamwork among students (c.f. Sect.~\ref{sec:teamwork}), and how others have tried to achieve a balance between technical and professional skills (c.f. Sect~\ref{sec:balance}). Finally, we give an overview of other experiences in SE course design (c.f.  Sect.~\ref{sec:courses}). 

\subsection{Enforcing teamwork among students}
\label{sec:teamwork}

Borrego et al.~\cite{borrego2013team} carried out a literature review regarding teamwork in engineering student projects. Many primary studies selected in this work report experiences from SE courses. According to this work, the major goals of team projects in engineering programs should be: encourage and promote teamwork, designing innovative and creative solutions taking into account ethics and efficiency, and develop communication skills, all this in a setting similar to what they should experience in industry once graduated. In order to motivate and retain students, programs should focus on real world problems. This is all in sync with the learning outcomes of CC4401. 

CC4401 can be considered a hybrid course, with elements of both project-based learning~\cite{blumenfeld1991motivating} and problem-based learning~\cite{barrett2010problem}. The course project is divided into 3-4 iterations, where each iteration can be considered a ``problem'', since there is a clear problem statement and goals which students must achieve, delivering a ``solution'' (requirements, user interfaces, source code, test cases and documentation). According to Blumenfeld et al.~\cite{blumenfeld1991motivating}, project-based learning is a comprehensive perspective focused on teaching by engaging students in research activities. Within this framework, students pursue solutions to non-trivial problems by asking and refining research questions, debating ideas, making predictions, designing plans and/or experiments, collecting and analyzing data, drawing conclusions, communicating their ideas and findings to others, asking new questions and creating artifacts. Project-based learning also places students in realistic, contextualized problem solving environments. In so doing, projects can serve to build bridges between phenomena in the classroom and real life experiences; the questions and answers that arise in the students' daily work gain value and are shown to be open to systematic inquiry.

According to Barrett~\cite{barrett2010problem}, under the problem-based learning approach, students are divided into teams and are presented with a problem they need to solve. Each group studies the problem, discusses and identifies what they should do in order to solve the problem. In the case of CC4401, teams focus on a different aspect of the course project during each iteration, building a ``mini-project'' solution, like designing adequate user interfaces. At the end of each iteration, the students present their solutions to the rest of course, and all teams discuss the solutions presented and must reach a consensus of which ``solution'' best solves the problem at hand.

\subsection{Achieving a balance between teaching technical and professional skills}
\label{sec:balance}

Schlimmer et al.~\cite{schlimmer1994team} mention that new graduates of SE programs \emph{``are also frequently unaware of current software practices and have difficulty fitting into an organized framework of software production. They need a more even balance between practical and theoretical knowledge.''} Students are not used to working in teams, so they are not always good at expressing their ideas, be it in writing or orally. They may also have attitudes that hinder effective communication. The authors conclude that undergraduate educators should make it their goal to make sure that students a conceptual and practical understanding of SE, as well as develop interpersonal working skills. As such, any SE curriculum should include project-based classes, providing an adequate combination of theory and practice. 

Soska et al.~\cite{soska2014implementation} analyzed job profiles for SE regarding cross-disciplinary competences, and the most important item that came up in the job posts is the ability to work in a team, with 53\% of the advertisements listing teamwork as needed. Furthermore, almost 30\% of the ads required that software engineers possess group communication and coordination skills. In the rest of this paper, a team is a group that \emph{``performs a defined, specialized task within a definite period, and whose members are generally cross-functional and disband after project termination''}~\cite{chiocchio2009cohesion}. Roshandel et al.~\cite{roshandel2011using} state that many SE courses now include  projects: team-based SE projects that students must develop using  techniques learned in classroom. 

\subsection{Other introductory SE teaching experiences}
\label{sec:courses}

During the first SE course, students usually work on tasks or small projects that look to solve only one part of a SE problem~\cite{garcia2004software}. These are usually either design tasks/projects or requirements elicitation and specification tasks. Rajlich~\cite{rajlich2013teaching} presented the introductory SE course that he teaches at Wayne University, where students work in teams on open source projects selected by the instructor, who also acts as a guide for these projects. Broman et al.~\cite{broman2010should} report that at Linköping University, the focus of the introductory SE course is on the software development process, where students develop small software projects that are not technically challenging. Drappa and Ludewig~\cite{drappa2000simulation} report the use of a simulation ``game'' to teach students the reality of SE. Bull and Whittle~\cite{bull2014observations} report that at Lancaster University, the introductory SE course is offered as a dedicated ``studio'' course for its students, the goal is to make students dive into the project and make the studio their second home.

\section{Redesigning CC4401}
\label{sec:intervention}

In this section, we first describe the original CC4401 course and the issues that we identified (c.f. Sect.~\ref{sec:original}). We then describe the new version of the course, explaining the mechanisms we added/changed in order to address the previously identified issues (c.f. Sect.~\ref{sec:new_version}).

\subsection{The original version of the course}
\label{sec:original}

\begin{table}[t]
    \centering
    \caption{Original course program, 2014.}
    \label{tab:original}
    \scalebox{0.75}{
\begin{tabular}{|r|p{3.3cm}|p{3.3cm}|}
\hline 
W1 & Introduction & The Nature of Software \\
W2 & Principles of SE & \textbf{Demo 0}  \\
W3 & No lecture & Requirements \\
W4 & Software Evolution & Student strike \\
W5 & Concept Location & Impact Analysis \\
W6 & Change Propagation & Review \\
W7 & \textbf{Evaluation 1} & \textbf{Demo 1} \\
\hline
 \multicolumn{3}{|c|}{Week-long student vacation} \\
\hline
W8 & Refactoring & Analysis Techniques \\
W9 & Testing Techniques & Modularization \\
W10 & Design Patterns & Design Patterns \\
W11 & Architecture & \textbf{Demo 2} \\
W12 & Review & \textbf{Evaluation 2} \\
W13 & SW Processes & SE in Practice \\
\cline{2-3}
W14 & \multicolumn{2}{c|}{No lectures, teams work on projects}\\
\cline{2-3}
W15 & No lecture & \textbf{Demo 3} \\
\hline 
\end{tabular}}
\end{table}

Table~\ref{tab:original} shows the program for the original version of CC4401. The first two weeks give students an overview of why SE emerged as a discipline, why building quality software is hard, and the principles that we follow as a discipline. We also explain the course project during the first lecture. Students must organize themselves into teams of 6-8 people, and by the end of the second week (\textbf{Demo 0}), each team must present the open source project that they plan to modify. This project must be an open source Java desktop application with 1-10 million LOC (our students are most familiar with Java). Students modified projects like the BlueJ and JEdit\footnote{\url{http://www.jedit.org/}} editors. 

Each team define candidate change requests for their project during \textbf{Demo 0} (bug fixes or new features), these are revised by the teaching staff. Each team appoints a coordinator for each iteration, who is in charge of making sure that the team will meet their obligations for the iteration. Over the next 4 weeks (W3-W6), students learn about the requirements engineering, as well as the software evolution process as formalized by Rajlich~\cite{Rajlich:2011}. During the first iteration, teams formalize their change requests as requirements, using concept location to identify where changes must be made, and evaluating the impact of the planned changes (\textbf{Demo 1}). 

During W7, students sat for an individual written evaluation, testing their knowledge of the SE concepts and techniques discussed in the lectures. Teams also present their first demo. We evaluated teams on how well they extracted requirements from their change requests (20\% of the demo grade), the quality of issues that they opened on Github (15\%), how orderly and systematic their concept location (20\%) and impact analysis (20\%) processes were. Finally, we also graded them on the quality of their oral presentations (10\%), their use of Github (10\%, to encourage frequent commits), and whether or not they defined a team coordinator (5\%). 

Teams continued working on their projects during weeks 8-11, and in lectures we discussed topics like refactoring, code inspections, testing, design and modularization, design  and architectural patterns. During \textbf{Demo 2}, teams give a live demo of the issues they have been working on, explaining which issues have been closed and which ones remain open. This demo is graded differently from \textbf{Demo 1}. 30\%  of the demo grade is an individual grade based on Github use. Another 30\% depends on how well the team used Github: were the individual branches merged back to the master branch? was the project adequately documented? The final 40\% of the grade depends on the quality of the live demo and slides. We also asked students to assess their teammates work habits and attitudes. These peer-evaluations were aggregated and sent to the students (without names), to help them identify how they could improve their teamwork. 

During W12, students again sit for a written evaluation, now about the concepts discussed during W8-11. We wrap up the lectures with a discussion about software development processes and SE in practice. The course ends with \textbf{Demo 3}, which is quite similar to \textbf{Demo 2}. The week after the semester ends, students can sit for a written exam  (mandatory for students with less than an A- average in the written evaluations). In order to pass the course, students must pass both the written evaluations and the project demos. 

A teaching assistant (TA) held weekly discussion sessions (1.5 hrs), half of this time was used to answer questions about the lecture material, the rest of the time was used to show students how to use the tools required  by the course (mainly Git and the JRipples~\footnote{\url{http://jripples.sourceforge.net/}} plugin for Eclipse). Students are expected to spend 10 hrs per week on this course. Two lectures and a discussion session meant that students have 5.5 hrs per week to work on their projects, prepare the demos and study for the written evaluations.

We detected several issues with this version of the course:
\begin{description}
\item [$I_1:$] Students felt that there was a certain disconnection between  the lectures and their projects, especially after \textbf{Demo 1}. 
\item [$I_2:$] Since the projects were large, some students felt perpetually ``lost'' and this affected their level of motivation. 
\item [$I_3:$] Some teams became groups, with students breaking the project up into chunks that they could handle by themselves. 
\item [$I_4:$] Some students felt that they were just coding to get a grade, and did little to improve their professional skills. For example, some teams delegated these tasks to those that they knew could do them more efficiently. 
\item [$I_5:$] The TA took too much time to give teams feedback. 
\item [$I_6:$] Feedback about presentations was given orally, and no detailed  feedback was given about the project documentation. 
\item [$I_7:$] Students could not switch teams when there were personal conflicts, since the learning curve for each project got steeper as the semester advanced. 
\end{description}

\subsection{The redesigned version}
\label{sec:new_version}

\begin{table}[t]
    \centering
    \caption{New course program, 2017 (first semester).}
    \label{tab:new_version}
    \scalebox{0.75}{
\begin{tabular}{|r|p{3.5cm}|p{3.5cm}|}
\hline 
W1 & Introduction & SW Quality \\
W2 & Requirements & Requirements  \\
W3 & SW Processes & SW Processes \\
W4 & SW Design & User Interaction \\
W5 & Wireframes \& Prototypes & Usability \\
W6 & Architecture & Architectural Patterns \\
W7 & Modularization & \textbf{Demo 1} \\
W8 & Review & \textbf{Evaluation 1} \\
W9 & Maintenance \& Evolution & Maintenance \& Evolution  \\
\hline
 \multicolumn{3}{|c|}{Week-long student vacation} \\
\hline
W10 & \multicolumn{2}{c|}{No lectures, teams work on projects}\\
\cline{2-3}
W11 & \textbf{Demo 2} & \textbf{Demo 2}  \\
W12 & Analysis Techniques & Functional Testing  \\
W13 & Testing Techniques & OO Testing \\
W14 & Review & \textbf{Evaluation 2} \\
W15 & \textbf{Demo 3} & \textbf{Demo 3} \\
\hline 
\end{tabular}}
\end{table}

Table~\ref{tab:new_version} shows the program for the new version of CC4401. Most of the topics are the same, albeit in a different order and depth. For example, the 4 lectures on software evolution (originally W4-6), have been reduced to two lectures during W9. We merged the second and third lectures, focusing on software quality. We also removed the lecture on Refactoring, since it is covered in depth in a pre-requisite course. We added extra lectures about requirements, software processes, user interaction, wireframes and prototypes, usability, and testing. With respect to the course project, we divided it into 4 iterations, where the last three have a live demo. The week-long student vacation is scheduled to coincide with national holidays, which is why it falls on a different week.

Teams no longer define their projects. Instead, the teaching staff presents a problem statement that the entire course will work on. We pick problems that we believe will motivate most students, like building a system to help them get lunch during the campus lunch rush-hour. Students are expected to build a simple web-based information system, allowing different types of users to edit, list and filter different elements. We also try to include API and library use, e.g., by asking them to geolocate data. The problem statement is intentionally vague, so that teams need to use the elicitation techniques discussed in class. Each team has up until W3 to pose up to 5 questions about the problem statement to the teaching staff, we use a shared Google Drive document. Each question is given a unique identifier, and we register which team asked the question.

The teaching staff answers these questions role-playing as users, usually once a day during W2-3, based on the idea of the system that we want students to develop. We try to be as precise as possible when answering, but will intentionally give vague answers when the questions are too open. We also sometimes tell students that we do not understand their questions, instead of answering what we think they tried to ask. This is because our students can get very technical, asking users how would they like a certain feature to be implemented. Once the deadline for the Q\&A part of this activity has passed, we publish a curated version of this document on the course website, this is the only source of information that students can use to define project requirements. Teams then have until the beginning of W4 to hand in a requirements specification document. To avoid ``busywork'', we do not ask teams to write a full requirements specification, instead asking them to define 10 functional and 3 non-functional requirements. They must clearly indicate which Q\&As were the source for these requirements, as well as how they would test them once the system is implemented. 

There are 3 live demos during the semester. Right after the students hand in their requirements, we give them our list of requirements, which we use to guide the development process. The list is longer than what the students should be able to implement during the semester, because we want students to get used to the idea that software development is an open-ended activity and that there is no clear ``end'' to a project. However, we are careful to clearly specify which requirements should be considered for each iteration. 

During \textbf{Demo 1}, teams showcase the user interface wireframes and/or prototypes that they developed to address certain requirements. Teams are free to use whatever technology they please during this phase. We do this to make them consider the different technologies seen in class, it is their responsibility to pick the best fit for their team (capability and time-wise). The only restriction is that they have to design interfaces for both desktop and mobile browsers. At the end of \textbf{Demo 1}, there is a vote. Students must individually vote for the set of interfaces that they believe the whole course should implement. The team that wins gets a bonus point for the next demo. 

We then specify the requirements for the next iteration, teams must implement the back-end for these requirements, connecting it to the provided user interfaces. This includes implementing mock-ups according to the look \& feel defined by the winning team. During \textbf{Demo 2}, teams must discuss the data model that they developed and demo their software. However, since we are now starting with smaller teams (3-4 people) and have a higher number of students, we schedule two demo days. This increase in students was expected, as students that are up to date with their courses  take CC4401 during the first semester of the corresponding year. All teams must hand in their slides the night before the first demos, and tag their code, thus no team has more time to prepare. Students must again vote, now picking the code base that will be used by all the teams during the final iteration. 

During the final iteration, most teams are now working with code that they did not developed (only the user interfaces remain unchanged), so we first ask them to develop a testing plan for the code, so as help them familiarize themselves with it and identify bugs. We then ask the teams to fix any bugs they find and finish implementing incomplete features. We may  ask teams to implement new requirements, depending on how buggy the  code is. 

The tight coupling between lectures and the project iterations helped us address issue $I_1$. During the first two iterations (requirements and user interfaces), teams start from scratch and get to pick the tools they want to use, which helps address $I_2$. Also, the TA covers the basics of web development and Django during the weekly discussion sessions. In order to address $I_3$, the TA helps teams set up and assign issues on Github, and we check the teams' Github activity when grading. There is still some ``slicing'' and ``chunking'' going on, but much less than before.  

To address issue $I_4$, teams must hand in technical reports for \textbf{Demos 1-3}, these are marked according to a rubric that takes into consideration both content and form. A well-formatted, -written and -structured report with the required content will get the top marks; one with the required content but with formatting/spelling errors will receive less marks, etc. The same goes for presentations, where we randomly pick a team member that has not yet presented. The use of rubrics also let us address issues $I_5$ and $I_6$. For starters, the iteration grades are usually ready within a week. However, we only publish these grades once the course has selected a winning team -- more on this in Sect.~\ref{sec:lessons}! The rubrics are published at the beginning of the corresponding iteration, so students know what to expect. Comments about the content, structure, format and writing style of the reports are given in writing. This feedback allows teams to better prepare the following reports and presentations. 

For each iteration, the report and presentation are worth 60\% and 30\% of the grade, respectively. The final 10\% corresponds to team peer-evaluations, where students use a 5-point Likert scale to grade their teammates on: their commitment to the project, whether they achieve their assigned tasks, whether they show initiative during development, whether they communicate and coordinate their work with the rest of the team, whether they hand in quality work, and if they are able to admit to mistakes and receive constructive criticism. They can (optionally) indicate positive and negative things about their teammates' behavior. All this is done using a web application, each student can see their own evaluation, all data is presented anonymously.

These peer-evaluations also let us know when there were problems within a team that, if left unattended, would affect the quality of the project, or even lead to failure. When we could not resolve an issue between team members, we opted to switch students between teams, ending up with teams of 4-5 people during the final iteration. This was possible since all the students are all working on the same code base, thus addressing issue $I_7$. We did not use the Github activity to grade students or teams during 2017, instead using it to corroborate situations where the team reported that someone did not do their share of the work. 

There are three slight differences between the 1st and 2nd semester of 2017. First, during the 1st semester, we let the students pick their own teams. In practice what happened was that: a) friends quickly teamed up, b) those without friends but that are more outgoing, quickly asked around and found suitable teams, and c) the quieter students had to make due with what was left. Since it was not clear if teams with these characteristics will actually work on their teamwork, we decided to create the teams during the second semester. We asked students to answer a simple diagnostic test, with 5 questions about their level of knowledge about the technology that they will use to develop the project, and 12 questions about their teamwork capabilities. We then used these scores to make sure that the initial teams were balanced with respect to technical knowledge, as well as teamwork capabilities. We also used these scores to make decisions when switching people between teams.

The second difference is that during the 1st semester, we did not ask students to do peer-evaluations during the requirements iteration. We later realized that this was a missed opportunity to give the students feedback, and since then we have included peer-evaluations in all iterations. Finally, the last difference between the two semesters is that we added a small, individual assignment at the beginning of the 2nd semester. Students were asked to build a simple to-do list using Django, due during W2. This was to ensure that students attended the first TA session on Django, and used the technology required by the course project in an early fashion.

\begin{table}[t]
\centering
\caption{Student demographic data.}
\label{tab:demografics}
    \scalebox{0.75}{

\begin{tabular}{l|c|c|c|l}
\cline{2-4}
& \multicolumn{1}{l|}{2014, 2nd sem.} & \multicolumn{1}{l|}{2017, 1st sem.} & \multicolumn{1}{l|}{2017, 2nd sem.} &  \\ 
\cline{1-4}

\multicolumn{1}{|l|}{\# Students} & 21                         & 40                               & 28                               &  \\ \cline{1-4}
\multicolumn{1}{|l|}{\# Female students}             & 2                         & 7                                & 4                                &  \\ \cline{1-4}
\multicolumn{1}{|l|}{\# Teams (start)}             & 3                         & 12                                & 9                                &  \\ \cline{1-4}
\multicolumn{1}{|l|}{\# Teams (end)}             & 3                         & 9                                & 7                                &  \\ \cline{1-4}
\multicolumn{1}{|l|}{Avg. project grade}             & 5.400                        & 6.1237                                & 6.1340                                &  \\ \cline{1-4}
\multicolumn{1}{|l|}{Avg. eval. grade}             & 5.0095                        & 5.3575                                & 5.5143                                &  \\ \cline{1-4}
\multicolumn{1}{|l|}{Avg. final grade}             & 5.1750                        & 5.8538                                & 5.7407                                &  \\ \cline{1-4}
\multicolumn{1}{|l|}{SD final grade}             & 1.2045                         & 0.5234                                & 0.70567                                &  \\ \cline{1-4}
\end{tabular}}
\end{table}

\begin{table*}
\centering
\caption{Student evaluations - 2017, 1st sem.}
\label{evaluations-autumn-2017}
    \scalebox{0.75}{
\begin{tabular}{l|c|c|c|c|c|c|c|c|c|c|c|cc}
\cline{2-12}
                                                                 & \multicolumn{2}{c|}{Iteration 1} & \multicolumn{3}{c|}{Iteration 2}        & \multicolumn{3}{c|}{Iteration 3}        & \multicolumn{3}{c|}{Iteration  4}       & \multicolumn{1}{l}{}              & \multicolumn{1}{l}{}              \\ \cline{2-14} 
\multicolumn{1}{c|}{}                                            & Report     & PE    & Report & PE & Pres. & Report & PE & Pres. & Report & PE & Pres. & \multicolumn{1}{c|}{Eval. 1} & \multicolumn{1}{c|}{Eval. 2} \\ \hline
\multicolumn{1}{|l|}{General}                                    
& 6.3000        & -                  & 6.5125    & 5.8450             & 6.7375          & 6.3000    & 5.8875             & 5.9550          & 5.8128    & 5.2692             & 6.5000          & \multicolumn{1}{c|}{5.0375}          & \multicolumn{1}{c|}{5.7436}          \\ \hline
\multicolumn{1}{|l|}{Teams with women (TW)}                           
& 6.2533        & -                  & 6.5800    & 6.6833             & 5.7200          & 6.2800    & 6.0917             & 5.7400          & 6.2333    & 6.5208             & 5.6667          & \multicolumn{1}{c|}{5.1700}          & \multicolumn{1}{c|}{5.9333}          \\ \hline
\multicolumn{1}{|l|}{Teams without women (TWW)}                        
& 6.2479        & -                  & 6.4857    & 6.7679             & 5.8714          & 6.3714    & 5.8786             & 6.1286          & 4.8917    & 6.4167            & 4.2917          & \multicolumn{1}{c|}{4.9750}          & \multicolumn{1}{c|}{5.5964}          \\ \hline
\multicolumn{1}{|l|}{Diff. \% TW and TWW} 
& 0.0867\%      & -                  & 1.4537\%  & -1.2489\%          & -2.5791\%       & -1.4350\% & 3.6249\%           & -6.3403\%       & 27.4276\% & 1.6234\%           & 32.0388\%       & \multicolumn{1}{c|}{3.9196\%}        & \multicolumn{1}{c|}{6.0200\%}        \\ \hline
\end{tabular}}
\end{table*}

\begin{table*}
\centering
\caption{Student evaluations - 2017, 2nd sem.}
\label{evaluations-spring-2017}
    \scalebox{0.75}{
\begin{tabular}{l|c|c|c|c|c|c|c|c|c|c|c|cc}
\cline{2-12}
& \multicolumn{2}{c|}{Iteration 1} & \multicolumn{3}{c|}{Iteration 2}        & \multicolumn{3}{c|}{Iteration 3}        & \multicolumn{3}{c|}{Iteration  4}       & \multicolumn{1}{l}{}              & \multicolumn{1}{l}{}              \\ \cline{2-14} 
\multicolumn{1}{c|}{}                                            & Report     & PE    & Report & PE & Pres. & Report & PE & Pres. & Report & PE & Pres. & \multicolumn{1}{c|}{Eval. 1} & \multicolumn{1}{c|}{Eval. 2} \\ \hline
\multicolumn{1}{|l|}{General}                                    
& 6.1036        & 6.2846                & 6.6536    & 5.6536             & 6.1643          & 5.6071    & 5.2393             & 6.3040          & 6.1786    & 5.5964             & 5.4964          & \multicolumn{1}{c|}{5.6392}          & \multicolumn{1}{c|}{5.4778}          \\ \hline
\multicolumn{1}{|l|}{Teams with women (TW)}                           
& 6.3000        & 6.6146                & 7.0000    & 6.8000             & 6.0750          & 5.5938    & 6.4750             & 5.4750          & 5.2688    & 6.5750            & 5.6188          & \multicolumn{1}{c|}{5.6125}          & \multicolumn{1}{c|}{5.5979}          \\ \hline
\multicolumn{1}{|l|}{Teams without women (TWW)}                        
& 5.8167        & 5.9931                & 6.2200    & 5.7633             & 5.2400          & 5.6150    & 6.2600             & 5.0550          & 5.8000    & 5.6500            & 5.5667          & \multicolumn{1}{c|}{5.6333}          & \multicolumn{1}{c|}{5.3600}          \\ \hline
\multicolumn{1}{|l|}{Diff. \% TW and TWW} 
& 8.3095\%      & 10.3708\%             & 12.5402\% & 17.9873\%          & 15.9351\%       & -0.3785\% & 3.4345\%           & 8.3086\%        & -9.1595\% & 16.3717\%          & 0.9356\%        & \multicolumn{1}{c|}{-0.3698\%}       & \multicolumn{1}{c|}{4.4387\%}        \\ \hline
\end{tabular}}
\end{table*}

\section{Results}
\label{sec:data}

Table~\ref{tab:demografics} shows data for three semesters of CC4401. First, we can see that there has been a slight increase in the number of female students between 2014 and 2017. Also, we did not change any of the student teams during 2014, starting and ending with 3 teams. During 2017, some teams changed up to two times (gained or lost a member), and we ended with less teams. Finally, rows 5-7 of Tab.~\ref{tab:demografics} show different averages: project, written evaluations, and final grades. All student grades are on a scale of 1.0 (min.) to 7.0 (max.), and 4.0 is a passing grade. The final grades for 2017 are slightly higher than those for 2014, and there is less variation in these grades. Note however that in 2014, the written examinations were worth more of the final grade (60\% instead of 50\%). We also used rubrics to evaluate the project artifacts during 2017, which may explain why there is less variation in these grades. We now analyze the data for 2017 in more depth. 

Since teams with women usually had at most one, we wanted to understand how they fared. We first look at the averages per iteration: for all teams (General), for teams with women (TW) and for teams without women (TWW) (c.f.  Tables~\ref{evaluations-autumn-2017} and~\ref{evaluations-spring-2017} for the 1st and 2nd semester of 2017, respectively). We did not include the final exam, since it is optional. Also, as explained in the previous section, there was no peer-evaluation (PE) during iteration 1 during the 1st semester. In the case of iterations 2-3, the report is worth 60\%, peer-evaluations 10\% and presentations 30\%. During the first semester, we see that the teams with women got lower presentation grades in iterations 2 and 3, and much higher grades during  iteration 4. We checked who presented each time: during iteration 2, three of the 5 teams with women had female presenters, during iteration 3 this number fell to 2 out of 5, and finally dropped to 1 out of 5 for the last iteration. We need to study this further: did these teams get better presentation grades at the end because less women presented? If so, what can we do to make the classroom a more welcoming space for women? On the other hand, we see that during the second semester, the teams with women always got better grades during presentations (iteration 2: 2 teams out of 4 had female presenters; iteration 3: 1 out of 4; iteration 4: 2 out of 4). 



We also studied how women fared in the peer-evaluations. Tables~\ref{peer-evaluation-autumn-2017} and~\ref{peer-evaluation-spring-2017} show the average peer-evaluation scores \emph{only} for teams that had at least one female team member (1st and 2nd semester of 2017, respectively). In general, women were better evaluated by their peers, except for team E during the 1st semester. This team can be considered an outlier: personal issues between these teammates meant that they could not continue working together, so we assigned each student to a separate team during the next iteration\footnote{All three students have already passed CC5401, the next SE course, where students develop a real system from scratch in bigger teams. So we would like to believe that they learned something about teamwork from their experience in CC4401. }.

\begin{table}
\centering
\caption{Comparison between female and male average peer-evaluations (APE-F and APE-M, respectively) for teams with women - 2017, 1st sem.
}
\label{peer-evaluation-autumn-2017}
    \scalebox{0.75}{
\begin{tabular}{|c|c|r|r|r|}
\hline
    Iter.                         & Team & APE-F & APE-M & Diff. \% APE-F and APE-M \\ \hline 
                               \multirow{5}{*}{\# 2} 
                                    & A    & 7.0000                         & 7.0000                         & 0.0000\%                                 \\  
                                                            & B    & 7.0000                         & 7.0000                         & 0.0000\%                                 \\  
                                                            & C    & 7.0000                         & 7.0000                         & 0.0000\%                                 \\ 
                                                            & D    & 7.0000                         & 6.7500                         & 3.7037\%                                 \\ 
                                                            & E    & 7.0000                         & 5.6667                         & 23.5294\%                                \\ \hline 
                               \multirow{5}{*}{\# 3} 
                               & A    & 6.4000                         & 5.5000                         & 16.3636\%                                \\  
                                                            & B    & 7.0000                         & 6.8000                         & 2.9412\%                                 \\  
                                                            & C    & 7.0000                         & 7.0000                         & 0.0000\%                                 \\  
                                                            & D    & 6.9000                         & 6.8250                         & 1.0989\%                                 \\  
                                                            & E    & 3.7000                         & 4.3333                         & -14.6154\%                               \\ \hline 
                               \multirow{5}{*}{\# 4} 
                               & A    & 6.9000                         & 6.5800                         & 4.8632\%                                 \\  
                                                            & B    & 7.0000                         & 7.0000                         & 0.0000\%                                 \\  
                                                            & C    & 5.9000                         & 6.2500                         & -5.6000\%                                \\  
                                                            & D    & 6.2000                         & 6.2200                         & -0.3215\%                                \\  
                                                            & E    & 7.0000                         & 6.2250                         & 12.4498\%                                \\ \hline
\end{tabular}}
\end{table}

\begin{table}
\centering
\caption{Comparison between female and male average peer-evaluations (APE-F and APE-M, respectively) for teams with women - 2017, 2nd sem.
}
\label{peer-evaluation-spring-2017}
    \scalebox{0.75}{
\begin{tabular}{|c|c|r|r|r|}
\hline
Iter. & Team & APE-F & APE-M & Diff. \% APE-F and APE-M \\ 
\hline 
 \multirow{4}{*}{\# 1} 
                            & A    & 6.5000                         & 6.1617                         & 5.4054\%                                 \\  
                              & B    & 6.6000                         & 6.6000                         & 0.000\%                                 \\  
                              & C    & 7.00                         & 6.9250                         & 1.0830\%                                 \\  
                              & D    & 7.0000                         & 6.7667                         & 3.4483\%                                 \\ \hline 
 \multirow{4}{*}{\# 2} 
                        & A    & 7.0000                         & 7.0000                         & 0.0000\%                                 \\  
                              & B    & 6.9000                         & 6.9667                         & -0.9569\%                                \\  
                              & C    & 7.0000                         & 7.0000                         & 0.0000\%                                 \\  
                              & D    & 6.4000                         & 6.2333                         & 2.6738\%                                 \\ \hline 
 \multirow{4}{*}{\# 3}  
                        & A    & 7.0000                         & 6.9000                         & 1.4493\%                                 \\  
                              & B    & 7.0000                         & 6.6000                         & 6.0606\%                                 \\  
                              & C    & 6.0000                         & 5.4000                         & 11.1111\%                                \\ 
                              & D    & 7.0000                         & 7.0000                         & 0.0000\%                                 \\ \hline 
 \multirow{4}{*}{\# 4}  
                            & A    & 6.9000                         & 6.9750                         & -1.0753\%                                \\ 
                              & B    & 5.5000                         & 7.0000                         & 27.2727\%                                \\ 
                              & C    & 6.9000                         & 6.8250                         & 1.0989\%                                 \\ 
                             & D    & 7.0000                         & 7.0000                         & 0.0000\%                                 \\ \hline 
\end{tabular}}
\end{table}

Each semester, we needed to reassigned some students to new teams and disbanded at least one team. This could affect student grades, especially during the 2nd semester, where we picked the initial teams. So, we analyzed the average grades for teams that did not change during the semester, and those that changed once or twice (c.f. Tables~\ref{rotation-autumn-2017} and~\ref{rotation-spring-2017} for the 1st and 2nd semester). No teams changed more than 2 times. To our surprise, there is no big difference between the teams that were stable during the whole semester and the teams were one or two changes occurred. 

\begin{table}
\centering
\caption{Iteration grades and team changes - 2017, 1st sem.}
\label{rotation-autumn-2017}
    \scalebox{0.75}{
\begin{tabular}{l|c|c|c|c|}
\cline{2-5}
                                            & Iter. 1 & Iter. 2 & Iter. 3 & Iter. 4 \\ \hline
\multicolumn{1}{|l|}{Teams without changes}  
& 6.3213         & 6.1583         & 5.9003         & 5.6969         \\ \hline
\multicolumn{1}{|l|}{Teams with 1 change} 
& 6.3083         & 6.3285         & 6.3366         & 5.9529         \\ \hline
\multicolumn{1}{|l|}{Team with 2 changes} 
& 6.2833         & 6.3317         & 6.0467         & 6.5350         \\ \hline
\multicolumn{1}{|l|}{General}               
& 6.3213         & 6.1583         & 5.90003         & 5.6969         \\ \hline
\end{tabular}}
\end{table}

\begin{table}
\centering
\caption{Iteration grades and team changes - 2017, 2nd sem.}
\label{rotation-spring-2017}
    \scalebox{0.75}{
\begin{tabular}{l|c|c|c|c|}
\cline{2-5}
                                            & Iter. 1 & Iter. 2 & Iter. 3 & Iter. 4 \\ \hline
\multicolumn{1}{|l|}{Teams without changes}  
& 6.1217         & 6.3046         & 5.5665         & 5.9357         \\ \hline
\multicolumn{1}{|l|}{Teams with 1 change} 
& 6.5094         & 6.2783         & 5.5963        & 5.5469         \\ \hline
\multicolumn{1}{|l|}{Team with 2 changes} 
& 5.7000         & 6.7267         & 5.7000         & 6.4450         \\ \hline
\multicolumn{1}{|l|}{General}               
& 6.1217         & 6.3046         & 5.5665         & 5.9357        \\ \hline
\end{tabular}}
\end{table}

\begin{figure}
\includegraphics[width=0.93\linewidth]{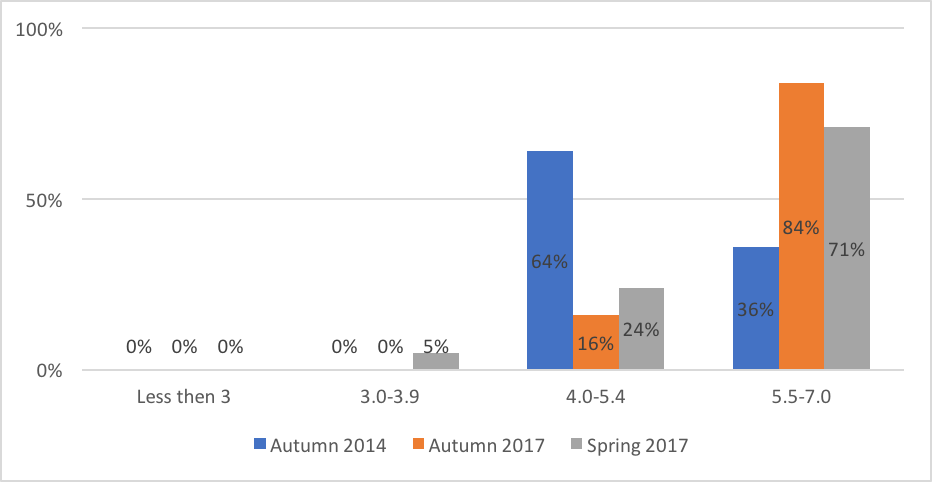}
\caption{How did you fare in the course?}
\label{fig1}
\end{figure}

\begin{figure}
\includegraphics[width=0.93\linewidth]{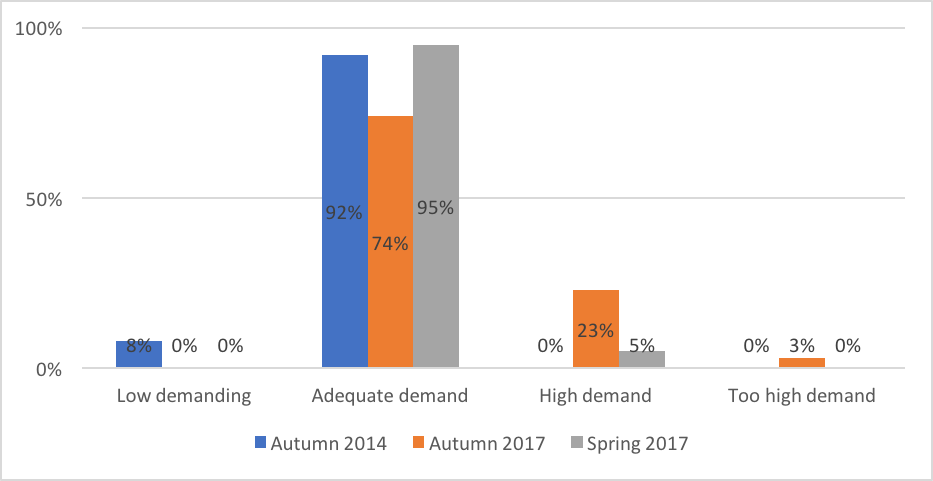}
\caption{How would you assess the course workload?}
\label{fig2}
\end{figure}

\begin{figure}
\includegraphics[width=0.93\linewidth]{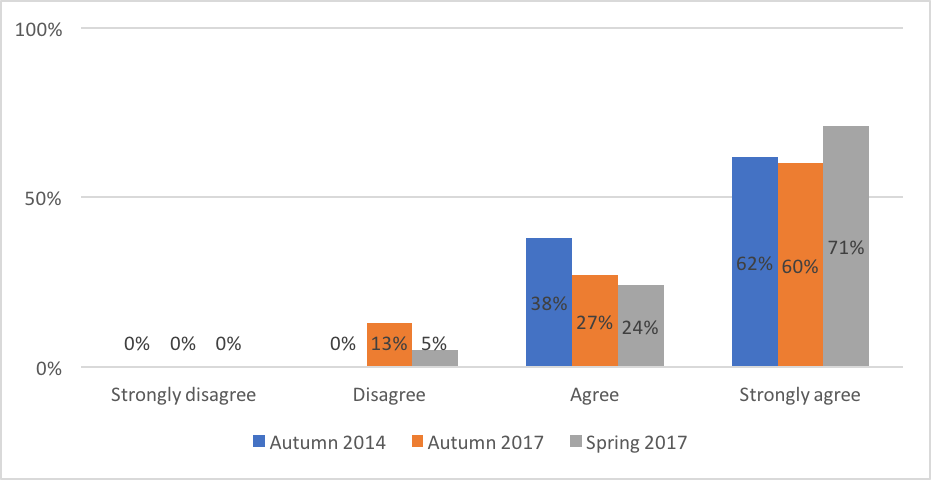}
\caption{The professor creates a space for reflection, posing relevant and challenging problems.}
\label{fig4}
\end{figure}

\begin{figure}
\includegraphics[width=0.93\linewidth]{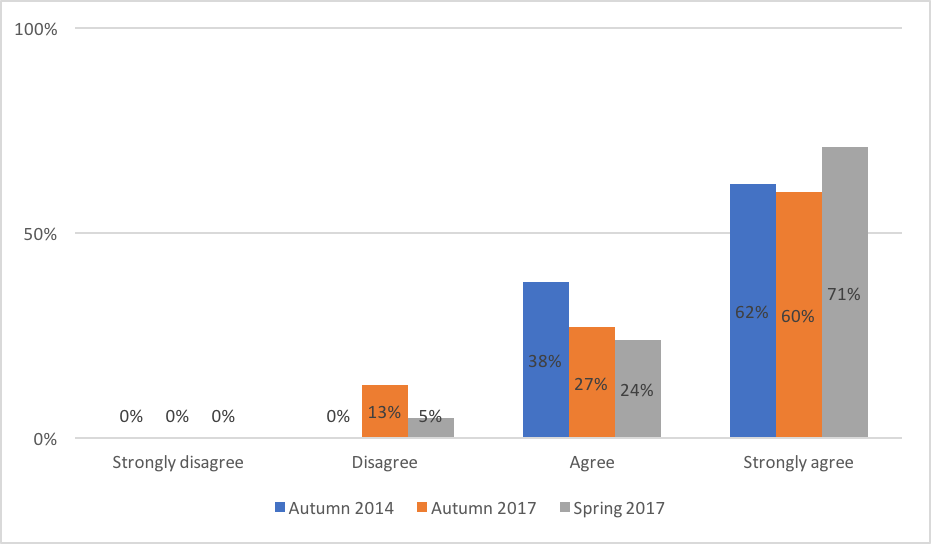}
\caption{The professor gives adequate and timely feedback about course evaluations. }
\label{fig5}
\end{figure}

At the end of each semester, our campus carries out student teaching evaluations. This survey is anonymous, and only consolidated data is made available. We now present the data for the 2014 course, as well as both 2017 semesters (all taught by the same instructor, the main author). We see in Fig.~\ref{fig1} that the new version of the course had a positive impact on the average grades, as perceived by the students\footnote{The 1st and 2nd semesters are ``Autumn'' and ``Spring'', respectively. }. In Fig.~\ref{fig2}, we see students' opinions on the perceived workload. During the 1st semester of 2017, students perceived a higher workload than 2014, but this has returned to normal levels with the changes we made last semester. Figure~\ref{fig4} shows that the strategy of picking a course project has been well-perceived by the students. Finally, in Fig.~\ref{fig5}, we see that students perceive that they receive slightly more feedback about evaluations, although there is still some room for improvement here.

Students can also give their opinion about the positive and negative aspects of the course (optional) in the survey. So what follows is anecdotal. During 2014, there were 4 positive comments, all about the teaching staff; and 5 negative comments, all about the course methodology. For example: \emph{``The lectures seem disconnected from the project after the first written evaluation. Since this is the first time we are working on software projects, it would be better to work on software that we already know, or that at least the TA knows''}. On the other hand, there were 11 positive comments after the 1st semester of 2017: 5 about the teaching staff and the rest about the course methodology. For example, one student said: \emph{``Apart from SE, I learned a lot about Django and web apps, which I have put to use in my personal projects. Before this, I knew practically nothing about web development. I also learned a lot about teamwork. It was a new experience to work with people that I did not know. It was also nice to put things into practice, given that there are few mandatory courses that do this, at least for now ... I loved the team changes during the last iteration (I'm not sure if this is because I saw new faces, or because I was now part of a larger group). I would consider doing this after each delivery. I think that the weight of the peer-evaluation was ideal; I would keep it that way''}. There were also 16 negative comments, most of them about how the TA discussion sessions were run and the contents of these sessions. Students expected a more guided introduction to Django and web technologies like Angular and React. After the 2nd semester of 2017, there were 8 positive and 6 negative comments. The positive ones were mostly about the course, e.g.: \emph{``This course is useful for LIFE. You could see that the professor really cared about how well we did, she adapted the course according to our needs. I liked that the project we developed can be of real use to society''} (they created a platform for reporting animal abuse). Again, the negative comments focused the TA sessions.

\section{Lessons Learned}
\label{sec:lessons}

In our opinion, this course teaches both theoretical and practical aspects of SE in a balanced manner, within a limited time frame (150 hrs). The dynamics of this course are quite changing: students have no guarantees that they will remain in the same team the whole semester, nor that they will continue working with their own code. In practice, students have to learn to get along and work with new people, toiling on code produced by others, which in all probability is buggy and incomplete. This is the nature of Millennials, changing, always getting challenged. 

The same goes for the teaching team: CC4401 is an ongoing learning experience for us. As an all-Millennial team, we chose to have an extremely flat team structure, meaning that we all took part in: designing the new teaching strategy, defining the course project and tools, make decisions about student teams, writing up evaluation questions and marking rubrics, as well as sharing part of the marking load. Also, since the course required some effort to run smoothly, we created a Telegram\footnote{Instant messaging application, similar to WhatsApp.} group to quickly share information about student teams, grades, etc., limiting the number of face-to-face meetings. We also stored all course documents on Google Drive, we all have full edition permissions. 

The social aspects of the course were also especially motivating for us as a teaching team. Not all software projects are money-oriented, a message that is sometimes lost in the discussion of project budgets, start-ups and large (rich) technology companies. The social slant that we gave the course project let us talk about social innovations in a very practical manner. This does require a certain amount of effort to set up, so for 2018 we have teamed up with Depachi, an on-campus student organization that focuses on developing software for non-profit organizations, our students will be developing a system for them. 

We feel that students learn more from the autonomy that we give them. This includes letting them pick what they consider to be the best ``solution'' for each iteration. This is why we only publish the grades for an iteration once a winning team has been picked -- we do not want them to pick the team that we thought was best, we want them to use the techniques they have learned in class to evaluate their classmates' solutions. Thus far, our students have owned up to their decisions, and have learned far more than we could have managed using more academic examples. 

For example, during the 1st semester of 2017, the team that won the iteration 3 vote had one of the most entertaining demos, since the team used memes and photos of the teaching staff. We had already discussed code reviews at this point in the lectures, so we were not expecting this team to win, as their code was a bit of a mess. They had implemented their own login system, since they decided that the one offered by Django was \emph{``too powerful for their needs''}. Most teams ended up refactoring the code so that they could use the Django login, since they quickly realized that the code they received was quite buggy! If we had designed this experience as part of the course, in all probability we would have had to bear the brunt of the students' frustration -- why would someone decide to reimplement such a fundamental component of the Django framework?! However, since their own classmates contributed this code, the teams focused on fixing it so that they could meet the requirements for the iteration. They also wanted to understand why someone would make an implementation decision like this. Students rarely ask questions during iteration presentations, in this case the team was bombarded with questions! 

During 2017, we only switched students between teams as a last resort, since we believed that the process would negatively affect student grades. However, it as been a pleasant surprise to see that this fear was unfounded, and that some students were very positive about this experience. We believe that the main reason for this is how we designed the project iterations: students must start with a new but somewhat familiar project each iteration. Even though we did not see any significant difference between letting students pick their teams and setting them up ourselves, we have decided to continue setting up the teams next semester. First, there is a question of fairness: when we pick the teams, they all start with a similar level of technical knowledge. Second, students are given an ``academic'' excuse to get to know new people, which is sometimes challenging for Millennials. Third, and last, the results of the diagnostic test can help figure out what needs to be covered by the TA in a more structured manner, which is an aspect of the course that still needs to be improved.

\bibliographystyle{ACM-Reference-Format-Journals}
\bibliography{acmtog-sample-bibfile}

\end{document}